\documentclass{article}
\usepackage{amsfonts}
\usepackage{amsmath}
\usepackage{epsfig}

\def\<{\left\langle}
\def\>{\right\rangle}

\begin{document}

\title{\hspace{4.1in}{\small CERN-PH-TH/2006-211} 
\\
\vspace*{1cm} \bf F-term uplifting via consistent D-terms}
\author{\vspace*{0.5cm} \bf Z. Lalak$^{a,b}$, O.J. Eyton-Williams$^{b}$, R. Matyszkiewicz$^{c,d}$ 
 \\
$^{a}$ Theory Division, CERN, 1211 Geneva 23, Switzerland\\
$^{b}$ Institute of Theoretical Physics, University of Warsaw, 
       00-681 Warsaw, Poland\\
$^{c}$ Institut f\"ur Theoretische Physik, TU Dresden, Germany\\
$^{d}$ Institute of Nuclear Physics, ul. Radzikowskiego 152, 31-342 Cracow, 
Poland}
\date{}
\maketitle
\centerline{}
\begin{abstract}
  The issue of fine-tuning necessary to achieve satisfactory degree of hierarchy between moduli masses, the gravitino mass and the
  scale of the cosmological constant has been revisited in the context of supergravities with consistent D-terms.  
We have studied (extended) racetrack
  models where supersymmetry breaking and moduli stabilisation cannot be separated from each other. We show that even in such
   cases the realistic hierarchy can be achieved on the expense of a single fine-tuning. The presence of two condensates changes the role of  the constant term in the superpotential,  $W_0$,
and solutions with small vacuum energy and large gravitino mass can be found even for very small 
values of $W_0$. Models where D-terms are allowed to vanish at finite vevs of moduli fields - denoted `cancellable' D-terms - 
and the ones where D-terms may vanish only at infinite vevs of some moduli - denoted `non-cancellable' - differ markedly in their properties. 
It turns out that the tuning with respect to the Planck scale required in the case of cancellable D-terms is much
  weaker than in the case of non-cancellable ones. 
  We have shown that, against intuition, a vanishing D-term can trigger F-term uplifting of the vacuum energy due to the stringent constraint it imposes on vacuum expectation values of charged fields.  Finally we note that our models only rely on two dimensionful parameters: $M_P$ and $W_0$.

\end{abstract}                                                                                                                      

\vskip 1cm

\section{Introduction}

The issue of hierarchical supersymmetry breakdown and moduli stabilisation returns as one of the leading themes in superstring
phenomenology. One needs to create a potential for various moduli fields which should simultaneously break supersymmetry,
split masses of superpartners in observable sector, and fix the remaining parameters of the low energy Lagrangian.  Selecting
the right vacuum is made more difficult by the existence of a trivial vacuum in stringy models, corresponding to a noninteracting
theory. To make the low-energy vacuum stable one needs to make the barrier between the finite and trivial vacua sufficiently high
and steep, which in particular implies that the masses of the moduli, like the dilaton and the volume modulus, should be notably
larger than the TeV scale (which is also the scale of the gravitino mass in scenarios with gravity mediation). In addition, the
scale of the vacuum energy has to be realistically small, orders of magnitude below the TeV scale.

The last requirement is particularly difficult to fulfil, as in generic models of spontaneous supergravity breaking the vacuum energy
tends to become negative, of the order $m_{3/2}^2 M_P^2$.

There are a limited number of options to uplift the vacuum energy to make it vanish or slightly positive. Firstly, one may add an
uplifting term which explicitly breaks local supersymmetry. This route has been known as the KKLT scenario, \cite{Kachru:2003aw,
  Kallosh:2004yh}, and has been widely explored by many authors. Secondly, the uplifting could be supersymmetric, if one finds a
way to cancel the negative definite term $-3 e^K |W|^2$ in the SUGRA scalar potential by a) $1/2 g^2 D^2$ -- D-term uplifting
or b) $e^ K K_{i \bar{j}} F^i \bar{F}^{\bar{j}}$ 
-- F-term uplifting. The first option has been explored in \cite{Achucarro:2006zf,Braun:2006se,Dudas:2005vv,Dudas:2006vc,Choi:2006bh}, and, in
\cite{Dudas:2005vv,Dudas:2006vc}, it was found that with a single
non-perturbative sector, the cancellation of the cosmological constant by the D-term  while keeping hierarchically small
gravitino mass is possible provided certain fine-tuning can be accepted and suitable corrections to the K\"ahler potential of the
volume modulus are introduced. Otherwise, as noted earlier in \cite{Achucarro:2006zf,Villadoro:2005yq,Villadoro:2006ia} one tends to obtain a heavy modulus and either a very heavy gravitino mass or an excessively large cosmological constant. 

An example of F-term cancellation is no-scale supergravity, which however doesn't stabilise the no-scale modulus at tree level. A
way out of this problem has been proposed in \cite{Lalak:2005hr} and further explored in \cite{Ellis:2006ar}, where a D-term was
suggested as an additional contribution fixing the value of the no-scale modulus. A general study of F-term uplifting given by
\cite{Gomez-Reino:2006dk} demonstrated that if supersymmetry is broken by F-terms from a sector that is not strongly influenced by
gravity, the SUSY breaking sector can act as an uplifting potential.  In fact, it has been subsequently shown that uplifting and
simultaneous stabilisation due solely to the F-terms is possible in simple cases of the O'Raifeartaigh type
\cite{Lebedev:2006qq}, with ISS-type \cite{Intriligator:2006dd} or Polonyi hidden sectors, considered in
\cite{Dudas:2006gr,Kallosh:2006dv,Abe:2006xp} and in warped backgrounds \cite{Brummer:2006dg}.  However, tuning is necessary in these models too, and in the supersymmetry breaking sector
one needs to give up on the (stringy) feature that all dynamically generated mass scales, which eventually provide explanation of
hierarchy between the Planck and the electroweak scales, are moduli dependent.

In this letter we present a refined mechanism of supersymmetric F-term
uplifting in the presence of consistent D-terms.  We introduce two
condensing sectors (thus effectively we are employing the racetrack
scheme). This has the advantage that the role of the constant term in
the superpotential, $W_0$, which is crucial in the schemes with a
single condensate, changes, as the two condensates compete with each
other and not with the $W_0$. In addition we introduce a spectator
charged scalar. As a result, we can rather easily, without any
dramatic tree-level hierarchy among the dimensionful parameters of the
Lagrangian, achieve nearly vanishing positive or negative (or zero)
cosmological constant with all moduli stabilised and small gravitino
mass. The D-terms vanish at the interesting minima, and the
cancellation of the negative $-3 e^K |W|^2$ is realised by the F-terms
of the moduli fields with the D-term acting indirectly, but
non-trivially, as a constraint on the F-term and superpotential
contributions to the potential. 
The D-term also serves as a source of large contribution to the mass
matrix, hence one of the fields becomes very heavy while the others
stay light. The existence of a state heavier than the condensation
scale is consistent with the fact that stabilisation and cancellation
of the vacuum energy result from the interplay between the gauge
sector and gravity-suppressed contributions.
A distinct feature of the model discussed here is that except $W_0$
which serves mainly as a source of mass for the lightest scalar, all
scales in the hidden sector are dynamically generated with all gauge
couplings explicitly field-dependent, as implied by string theory, in
contrast to models with ISS-type hidden sectors,
\cite{Intriligator:2006dd}, considered in
\cite{Dudas:2006gr,Kallosh:2006dv}, where the crucial mass scale in
the hidden sector has been put in arbitrarily.

\section{General setup}
To start with let us assume for simplicity that the dilaton becomes stabilized and decouples at some scale $M_s$ close to the
Planck scale, and below $M_s$ the superpotential for the light volume modulus takes the form $W=W_0 + W(T)$, where $W_0 = \langle
W(S) \rangle$.  With the simple expression $W(T)= A e^{-a T}$, one stabilizes the $T$-modulus but with a negative cosmological
constant, $V_0 < 0$.  This can be remedied by adding in a positive ``D-term'' contribution $\delta_D V$ which depends on $T$ in
such a way that a) it makes the $V_0 $ positive or zero, and b) it doesn't spoil the stabilization of $T$.  In the scenario presented
by KKLT it has been assumed that $\delta_D V = 1/2 g^2 D^2 $, where $D=1/(T + \bar{T})^n$. This can be seen as arising from
brane-antibrane forces, but then supersymmetry is explicitly broken in the effective 4d Lagrangian. This makes it difficult to
reliably study questions about soft supersymmetry breaking within 4d field theoretical framework \cite{Choi:2004sx}.

Alternatively, an interesting possibility is that the uplifting term could be 
a FI term coming from gauging of a shift symmetry \cite{Burgess:2003ic}.  However, as noticed by Dudas et al. \cite{Dudas:2005vv} the simple superpotential $W=W_0 -
A e^{-a T}$ is not invariant under the local imaginary shift of $T$: $T \rightarrow T + i \delta /2 \Lambda$ which produces
D-terms of the required form.  It turns out that introducing a gauge invariant version of the superpotential changes the rules of
the game. We shall follow this route below, putting aside the possibility of gauging an R-symmetry\footnote{{}For the R-symmetric scheme to work, all terms
in the superpotential must acquire the same local phase, hence the constant term $W_0$ is excluded, and many other useful terms
are problematic, like a purely dilatonic piece.  Nevertheless, it has been shown in that one can construct simple models of this
type which have flat or dS minima for $T$ with broken supersymmetry.}, see Zwirner et al.\cite{Villadoro:2006ia}, Binetruy
et al.\cite{Binetruy:2004hh}. 
  
\subsection{Gauging the shift}

In this letter we will limit our focus to models which are invariant under non-R $U(1)$ gauge symmetry.
Let us consider gauging of the imaginary shift of the modulus $T$. It has been shown in \cite{Haack:2006cy} that one can indeed realise
such a situation in stringy models. Let $K=K(T + \bar{T})$. The shift $T \rightarrow T + i \delta /2 \Lambda$ is generated by the
Killing vector $X_T = i \delta/2$ where $\delta $ is real. The prepotential $D$ fulfills the Killing equation $K_{T\bar{T}}
\bar{X} = i \partial D/ \partial T$, and generates the scalar potential $\delta V = 1/2 g^2 D^2 $.  One finds, upon solving the
Killing equation, $D=- \frac{\delta}{2} \frac{\partial K}{\partial \bar{T}} + \xi$, where $\xi$ is a genuine constant. If $\xi
\neq 0$ the gauged symmetry acts on gravitini and becomes a local $U(1)$ R-symmetry, as described earlier. We shall set $\xi=0$ in
what follows.  To be consistent with global supersymmetry algebra the superpotential must be invariant under any local (non-R)
symmetry.  The invariant superpotential which is solely a function of the modulus $T$ must therefore be a constant. This is not sufficient to
stabilize the $T$.

The solution is to introduce more charged fields:
\begin{equation}
\delta Z = i \Lambda q_Z Z; \;\; X_Z=i q_Z Z.
\end{equation}

\subsection{Gaugino condensation}
Let us assume that the source of the superpotential for $T$ is non-perturbative effects in a strongly coupled gauge theory. For
simplicity let us take $SU(N)$ hidden gauge group with $N_f < N$ pairs of $N$ and $\bar{N}$ representations. In addition, let us
assume that the fields are charged under a local $U(1)$. Below the condensation scale the effective non-perturbative superpotential
is
\begin{equation}
  W_{npert} = (N-N_f) \left ( \frac{\Lambda_{cond}^{b_0}}{det M} \right 
  )^{\frac{1}{N-N_f}},
\end{equation}
where $b_0=3 N -N_f$, $M_{ij}=Q_i \bar{Q}_j$. The condensation scale is related to the modulus $T$ through the relation
$\Lambda_{cond} = e^{-T/b_0} $ where $Re(T) =\frac{ 8 \pi^2}{g^2}$.  Let us take $N_f=1$. Then the superpotential becomes
\begin{equation}
  W_{npert} = (N-1) \left ( \frac{e^{-T}}{M} \right )^{\frac{1}{N-1}}. 
\end{equation}
This is invariant under the local $U(1)$ if $M \rightarrow M e^{i(q+\bar{q}) \Lambda}$ with $\delta = - 2 (q+\bar{q})$.  In
general this means that $Tr(Q_{U(1)})$ is nonzero, so naively it is anomalous. However, the 4d Green-Schwarz mechanism may be employed to cancel the
one-loop anomalous diagrams, or one can imagine adding to the model a number of charged fields which do not enter the superpotential.

To be able to write down the effective potential one needs also the K\"ahler function for the composite meson fields M.  Here one
should use symmetries. Possible choices are for instance $K=\bar{M}{M}/M_{P}^2$ or $K= \sqrt{ \bar{M}{M}}$.  Since we do not
believe that a particular choice of the K\"ahler potential could play a significant role, we take the first option which leads to
cleaner formulae.

\section{Racetrack with consistent D-terms}

The simplest model that works analogously to the original KKLT model is the one with a racetrack superpotential stabilizing $T$:
\begin{equation}
  W= A_1 N_1 \left ( \frac{e^{-T}}{M_1} \right )^{\frac{1}{N_1}} - A_2 N_2 \left ( 
    \frac{e^{-T}}{M_2} \right )^{\frac{1}{N_2}}+ W_0,
\end{equation}
with
\begin{equation}
  V_D = \frac{\pi^2 \delta^2}{t} (\frac{b}{2t} + x_{1}^2 + x_{2}^2
  )^2,
\end{equation}
where $x_i=| M_i | \geq 0$. We assume the K\"ahler function
\begin{equation}
  K=-b\log(T+\bar{T})+M_1\bar{M}_1+M_2\bar{M}_2\ ,
\end{equation}
with $b=1,2 \;{\rm or} \;3$. Notice that there could be a constant contribution to the K\"ahler potential\footnote{Such a constant may be a remnant left over after decoupling of very heavy fields, like the dilaton.}, 
$c_K$, which would
rescale uniformly all terms in the scalar potential by $e^{c_K}$, except the contribution $V_D$. In this respect
$c_K$ 
acts in the way analogous to that of the charge $\delta$, which in turn rescales {\em only} $V_D$.  The superpotential is
defined up to the overall complex phase. Hence, one can fix that phase in such a way that the constant part of the superpotential
($W_0$) can be considered to be a real number.  Then the scalar potential takes the form
\begin{equation}
V=V_F+V_W+V_D\ , \label{eq:V}
\end{equation}
with
\begin{align}
  V_F=(2t)^{-b} e^{x_1^2+x_2^2}\Bigg[ &|A_1|^2 \left(\frac{e^{-
        2t}}{x_1^2}\right)^{\frac{1}{N_1}}\left[b\left(\frac{2t}{b}+N_1\right)^2
    +x_1^2\left(N_1-\frac{1}{x_1^2}\right)^2+x_2^2N_1^2\right]\nonumber\\
  &+|A_2|^2 \left(\frac{e^{- 2t}}{x_2^2}\right)^{\frac{1}{N_2}}\left[b\left(\frac{2t}{b}+N_2\right)^2
    +x_2^2\left(N_2-\frac{1}{x_2^2}\right)^2+x_1^2 N_2^2\right]\nonumber\\
  &+W_0^2\left[b+x_1^2+x_2^2\right]\nonumber\\
  & \begin{aligned} -\ 2|A_1||A_2|\left(\frac{e^{- t}}{x_1}\right)^{\frac{1}{N_1}}\left(\frac{e^{-
          t}}{x_2}\right)^{\frac{1}{N_2}}\cos\left(\phi_1-\phi_2\right) \Bigg[&
    b\left(\frac{2t}{b}+N_1\right)\left(\frac{2t}{b}+N_2\right)\nonumber\\
    & +x_1^2\left(N_1-\frac{1}{x_1^2}\right)N_2+x_2^2\left(N_2- \frac{1}{x_2^2}\right)N_1\Bigg]
    \end{aligned} \nonumber\\
    & +2W_0|A_1|
    \left(\frac{e^{-t}}{x_1}\right)^{\frac{1}{N_1}}\cos\left(\phi_1\right)
    \left[2t+N_1 b+N_1 \left(x_1^2+x_2^2\right)-1\right]\nonumber\\
    &-2W_0|A_2|
    \left(\frac{e^{-t}}{x_2}\right)^{\frac{1}{N_2}}\cos\left(\phi_2\right)
    \left[2t+N_2 b+N_2 \left(x_1^2+x_2^2\right)-1\right]\Bigg]\ ,
\end{align}
and
\begin{align}
  V_W=-3(2t)^{-b} e^{x_1^2+x_2^2}\Bigg[ &|A_1|^2 \left(\frac{e^{-2t}}{x_1^2}\right)^{\frac{1}{N_1}}N_1^2+|A_2|^2
  \left(\frac{e^{-2t}}{x_2^2}\right)^{\frac{1}{N_2}}N_2^2+W_0^2\nonumber\\
  & -2|A_1||A_2|\left(\frac{e^{- t}}{x_1}\right)^{\frac{1}{N_1}}\left(\frac{e^{-
        t}}{x_2}\right)^{\frac{1}{N_2}}\cos\left(\phi_1-\phi_2\right)
  N_1 N_2\nonumber\\
  &+2W_0|A_1|\left(\frac{e^{- t}}{x_1}\right)^{\frac{1}{N_1}}\cos\left(\phi_1\right)N_1 -2W_0|A_2|
  \left(\frac{e^{-t}}{x_2}\right)^{\frac{1}{N_2}}\cos\left(\phi_2\right)N_2 \Bigg]\ ,
\end{align}
where $A_1=|A_1|e^{ i\omega_1}, \; A_2  =|A_2|e^{i\omega_2}, \; M_1=x_1e^{ i(\phi_1+\omega_1)N_1}, \; M_2 =x_2 e^{ i(\phi_2   +\omega_2)N_2}, \;
T=t+ i a \, $.

At first glance we have 6 degrees of freedom in this model: $t$, $x_1$, $x_2$, $a$, $\phi_1$ and $\phi_2$. In fact only 5 of them
are relevant: certain combination of $a$ and two phases $\phi_1$ and $\phi_2$ forms the longitudinal polarisation of the massive
gauge boson, that appears due to spontaneous breakdown of the shift symmetry. Hence, one can gauge away one combination of these
fields.  Alternatively, one may choose to work with a flat direction in the scalar
potential. 
Hence in practice one needs to solve the potential with respect to $t$, $x_1$, $x_2$, $\phi_1$, $\phi_2$ and $a$ then the flat
direction can be identified with the longitudinal polarisation.  In addition, different choices of $\omega_1$ and $\omega_2$
correspond to the different settings of the origins of the $\phi_1$ and $\phi_2$ variables respectively. Without lose of
generality, one can put $\omega_1=\omega_2=0$.

The behaviour of this model is close to that of the racetrack, with the $t$ modulus being stabilised by its F-term potential.  The
minimum appears close to\footnote{This expression is obtained by solving $F_T=0$ for t.}:
\begin{equation}
  \<t\>= \ln\left(\frac{|A_1| x_2^{1/N_2}}{|A_2| x_1^{1/N_1}}\right)\frac{N_1 N_2}{|N_2-N_1|} \label{eq:approx_t_vev}
\end{equation}
where the dynamical phases absorb any overall phases, such as the sign of $N_2-N_1$. We expect $x_1$ and $x_2$ to be typically
order one because they have no sources of hierarchy.  Hence the minimum reached will not be exactly the value given by
Eq.~(\ref{eq:approx_t_vev}) since there are additional contributions to the potential coming from the F-terms for the condensate
fields and the D-term.  However, for the values of the parameters considered in this paper the differences are minimal.

In addition to $t$, the $x_1$ and $x_2$ fields need stabilising. To see that this must happen we need only consider $V_W$ and
$V_F$.  Since $V_W$ only contains negative, fractional powers of $x_1$ and $x_2$ it is unbounded from below with $x_1$ and $x_2$
tending to zero.  However, $V_F$ includes larger negative powers with positive coefficients, coming from the $\frac{\partial W}{\partial x_i}$ terms, and the
negative singularity is lifted to be a positive singularity.  Finally, $x_1$ and $x_2$ cannot run off to infinity since the F-terms for $M_1$ and $M_2$
contain positive powers of $x_1$ and $x_2$ cancelling the negative powers coming from the superpotential.

Since the phases are cyclic, by definition, stability is guaranteed and the number of flat directions is determined by the global
symmetries.  With all input parameters set to non-zero values there exists one flat direction corresponding to the anomalous
$U(1)$ gauge symmetry.

At this stage, the potential is stable without needing to utilise the D-term.  It is clear that the D-term contribution alone
cannot stabilise $t$ since it runs off to $\infty$, but it does perturb the minimum reached.  The importance of the D-term is in
cancelling the cosmological constant.  Since $V_D$ is very flat at large $t$ its effect on the position of the minimum is
minor\footnote{Assuming that $\delta$ is small; if $\delta$ is too large $\frac{\partial V_F}{\partial t}$ cannot cancel
  $\frac{\partial V_D}{\partial t}$ and there is no solution to $\frac{\partial V}{\partial t}=0$.}, but it provides a positive
definite contribution to $V$, which allows the cosmological constant to be tuned to zero with arbitrary precision.  As expected this closely mirrors the behaviour in the original 
KKLT formulation, with the D-term lifting playing the role of the explicit SUSY
breaking anti--D-brane contribution.

In addition to the direct contribution of $V_D$'s energy density to $V$, the D-term indirectly increases the overall energy density through its effect on $V_F$ and $V_W$.  
Since $V_F+V_W$ minimise
with $\delta=0$ it follows that any alteration of the position of the minimum will increase the value of $V_F+V_W$ (assuming no
flat directions). It is the combination of these two effects that gives rise to the final expectation values minimising the potential.

Unfortunately these effects are too efficient.  A very small $\delta \sim 10^{-14}$ is sufficient to lift from a negative to a
positive minimum and a natural, order one, $\delta$ does not allow finite minimisation. In principle, from the point of view of the
effective field theory a very small effective $U(1)$ charge of $T$ and $M_i$ could be tolerable although it is highly unnatural, and it is also unnatural from the point of view of string theory. 
Thus we shall try to improve the model in such a way that solutions with $\delta\sim 1$ could be found. To this end we consider a cancellable D-term in the next section.

At this point it is instructive to have a closer look at the breaking of the anomalous $U(1)$.  The mass terms come from the kinetic
part of the Lagrangian
\begin{equation}
  \frac{b \delta^2}{16 t^2} (V_\mu - \frac{2}{\delta} \partial_\mu {\rm Im}(T) )^2 + 
  x_i^{2} q^{2}_{M_i} (V_\mu - \frac{1}{q_{M_i}} \partial_\mu 
  \phi_i N_i )^2,
\end{equation}
where $M_i = x_i  e^{i \phi_i N_i}$ and summation over charged fields is understood. One easily finds out that 
the mass of the vector boson is given by
\begin{equation}
  m^{2}_V = \frac{8\pi^2}{t}\left(\frac{b \delta^2}{8 t^2} + 2 q^{2}_{M_i} x_i^{2}\right)=\frac{4\pi^2\delta^2}{t} (\frac{b}{4 t^2} + x_i^2).
\end{equation}
It turns out that in the models we consider here the `anomalous' gauge
boson is always rather heavy, see table~1.

\section{Natural D-terms and natural charges}

Since the D-term in the previous model only minimises for $t\rightarrow\infty$, $x_1\rightarrow 0$ and $x_2\rightarrow 0$ and
contains no sources of hierarchy order one values of $t$, $x_1$ and $x_2$ imply that $V_D$ will be order one.  With $V_W \sim
10^{-32}$, to obtain a physically reasonable gravitino mass it is clear that $V_D$ does not obtain a natural vev, as demonstrated
by the smallness of $\delta$.  One approach to this problem is to introduce a field that appears with a relative minus sign inside
the D-term, such that $V_D$ can go to zero for natural input parameter values.

The effect of a cancellable D-term is purely the introduction of a constraint on the rest of the potential.  In the limit that the
solutions to $\frac{\partial V}{\partial Z_i}=0$, where $Z_i$ runs over all modulus fields, are governed by the first
derivatives of $V_D$ then the solution\footnote{There is only one non-trivial solution, in the limit $\delta\rightarrow\infty$,
  namely the solution in which the D-term goes to zero} to $\frac{\partial V_D}{\partial Z_i}=0$ can be imposed as a constraint
on $V_F$ and $V_W$ and, to good approximation, $V_D=0$.  The simplest possible D-term of this form is obtained by introducing one
additional matter field, $C$, with the opposite sign charge to the condensing fields.  It sufficient to consider an additional
field which does not appear in the superpotential, only in $V_D$ and $V_F$.  We choose to consider a canonical K\"ahler potential for simplicity. 

\subsection{Additional Charged Matter}

As $C$ does not appear in the superpotential it only interacts via the D-term and through the K{\"a}hler derivatives, in the
F-terms.  This leads to a change of $V_{W}\rightarrow e^{c^2}V_{W}$ and a change of $V_F \rightarrow e^{c^2}V_F + V_F(c)$, where $c=|C|$ and
\begin{align}
  V_F(c)=c^2e^{K}|W|^2=(2t)^{-b}c^2 e^{c^2} e^{x_1^2+x_2^2}\Bigg[ &|A_1|^2 \left(\frac{e^{-2t}}{x_1^2}\right)^{\frac{1}{N_1}}N_1^2+|A_2|^2
  \left(\frac{e^{-2t}}{x_2^2}\right)^{\frac{1}{N_2}}N_2^2+W_0^2\nonumber\\
  &
  -2|A_1||A_2|\left(\frac{e^{-t}}{x_1}\right)^{\frac{1}{N_1}}\left(\frac{e^{-t}}{x_2}\right)^{\frac{1}{N_2}}\cos\left(\phi_1-\phi_2\right)
  N_1 N_2\nonumber\\
  &+2W_0|A_1|\left(\frac{e^{-t}}{x_1}\right)^{\frac{1}{N_1}}\cos\left(\phi_1\right)N_1 -2W_0|A_2|
  \left(\frac{e^{-t}}{x_2}\right)^{\frac{1}{N_2}}\cos\left(\phi_2\right)N_2 \Bigg]\ .
\end{align}

The D-term potential is given by
\begin{align}
  V_D = \frac{\pi^2 \delta^2}{t} \left(\frac{b}{2t} + x_{1}^2 + x_{2}^2-q c^2\right)^2
\end{align}
where $q$ is the charge of $C$, defined such that $C\rightarrow e^{iq \delta \Lambda}C$ under the anomalous $U(1)$.  If $c$ had no
potential, besides the D-term, it would simply move such that it was given by
$c^2=\frac{1}{q}\left(\frac{b}{2t}+x_1^2+x_2^2\right)$ and the minimum for $V_F+V_W$ would be undisturbed.  However $c$ has a mass
term from its coupling to $|W|^2$ and as such is driven towards zero.  Exactly how much $c$ changes from the default value is
fixed by the interplay between the minimisation of $t$, $x_1$ and $x_2$ and the minimisation of $c^2|W|^2$.  The deviation is
determined by the size of the mass term for the constraint equation, which is in turn given by $\frac{1}{|q|}|W|^2$.  The smaller
$|q|$ is the larger the effective mass term for the constraint equation and hence the further the solution to the equation will be
driven from the preferred value of $V_F+V_W$.  Since moving away from this value will increase the vev of $V_F+V_W$ a small enough
charge lifts the cosmological constant to be positive\footnote{With $W_0=0$ and all other parameter values given by
  table~\ref{tab:inputData}, $q\sim1$ is sufficient.}.  In addition to this indirect lifting, the extra matter provides a positive
contribution to energy density through the mass term $c^2|W|^2$.  However, since $c^2|W|^2$ is driven to zero this effect is less
important than the effect on $V_F+V_W$ and, alone, is insufficient to cancel $V_W$.  The combined effects can lift the potential
to be small and positive, as shown in our numerical examples.

The cosmological constant can be sent to zero by tuning the input parameters.  It is clear from the previous discussion that there
exists a value of the charge $q$ that gives $V=0$, but perhaps a more natural parameter to tune is $W_0$.  Considering a small
$W_0$ we see that there is a negligible positive contribution proportional to $W_0^2$, and two terms coupled to cosines of
$\phi_1$ and $\phi_2$.  For small $W_0$ the $|A_1| |A_2| \cos(\phi_1 -\phi_2)$ term dominates and hence enforces $\phi_1=\phi_2$.
Nonetheless $\phi_1$ is still free to align such that the coefficient of $W_0$ is negative.  Thus a model with a small $W_0$ will
always result in a lower minimum than the same model with $W_0=0$.  As a result, $W_0$ can be tuned such that $V=0$, assuming that
a $q$ is chosen such that when $W_0=0$ the minimum is positive.

\section{Discussion}

We present numerical results and provide approximate, analytic justifications of the results seen.

\begin{table}[!ht]
  \centering
  \begin{math}
    \begin{array}{|c||c|c|}
      \hline
      \text{ Data} & \text{ Non-cancellable }  & \text{Cancellable}  \\
      \hline
      m_{3/2} & 227.761 & 478.142 \\
      \hline
      V_0^{1/4} & 0 & 0 \\
      \hline
      V_F^{1/4} & 2.13211 \times 10^{10} & 3.41239 \times 10^{10}  \\
      \hline
      V_D^{1/4} & 1.78276 \times 10^{10} & 2292.40 \\
      \hline
      t & 732.348 & 733.325\\ 
      \hline
      x_1 & 0.169201 & 0.173619 \\
      \hline
      x_2 & 0.208768 & 0.237810 \\
      \hline
      \phi_1 & \pi & \pi \\
      \hline
      \phi_2 & \pi & \pi \\
      \hline
      m_t & 8.213 \times 10^4 & 1.630 \times 10^5 \\
      \hline
      m_a & 8.277 \times 10^4 &  1.646 \times 10^5  \\
      \hline
      m_{x_1} & 1.760 \times 10^3 & 3.141 \times 10^2  \\
      \hline
      m_{x_2} & 2.369 \times 10^3 & 2.427 \times 10^{17}  \\
      \hline
      m_c & \text{N/A} & 7.853 \times 10^2 \\
      \hline
      m_{\phi_1} & 4.914 \times 10^2 & 6.000 \times 10^2  \\
      \hline
      m_{\phi_2} & 0 & 0 \\
      \hline
      m_V & 4.73 \times 10^3 & 8.56 \times 10^{17} \\
      \hline
    \end{array}
  \end{math}
  \caption{Comparison of the models.  The cosmological constant can be tuned to be as small as the numerical precision allows and hence is set to zero.}
  \label{tab:AllModels}
\end{table}
In the tables $V_0=\<V\>$ is the cosmological constant and all dimensionful quantities are given in GeV except for the field vevs which are
quoted in units in which $M_P=1$.  Finally, note that the masses quoted are not the masses of the fundamental fields, for which
the mass matrix is not diagonal.  Instead the masses are labelled by the fundamental field that makes up the largest component of
that mass eigenstate.

\begin{table}[!ht]
  \centering
  \begin{math}
    \begin{array}{|c||c|c|}
      \hline
      \text{ Data} & \text{ Non-cancellable }  & \text{ Cancellable}  \\
      \hline
      b & 3 & 3 \\
      \hline
      |A_1| & 2.62 & 2.62  \\
      \hline
      |A_2| & 0.3 & 0.3  \\
      \hline
      N_1 & 25 & 25  \\
      \hline
      N_2 & 27 & 27 \\
      \hline
      \delta & 6.22 \times 10^{-15} & 1  \\
      \hline
      W_0 & 2 \times 10^{-12} & 2.13 \times 10^{-12}  \\
      \hline
      q & N/A & 1/2 \\
      \hline
    \end{array}
  \end{math}
  \caption{Input parameter values.}
  \label{tab:inputData}
\end{table}

\begin{figure}[htp]
  \includegraphics{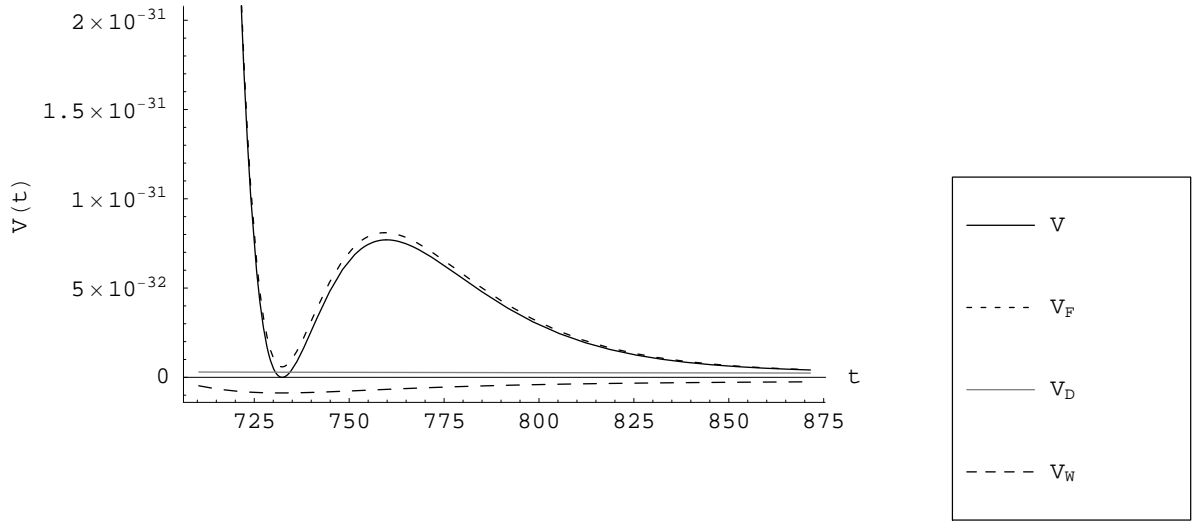}\caption{Minimisation with a non-cancellable D-term.  In addition to the total potential, the F-term, D-term and superpotential contributions are plotted.  All quantities are given in units where $M_P=1$ and all other fields are fixed at their minimum values.} \label{fig:noncan}
\end{figure}

\begin{figure}[htp]
  \includegraphics{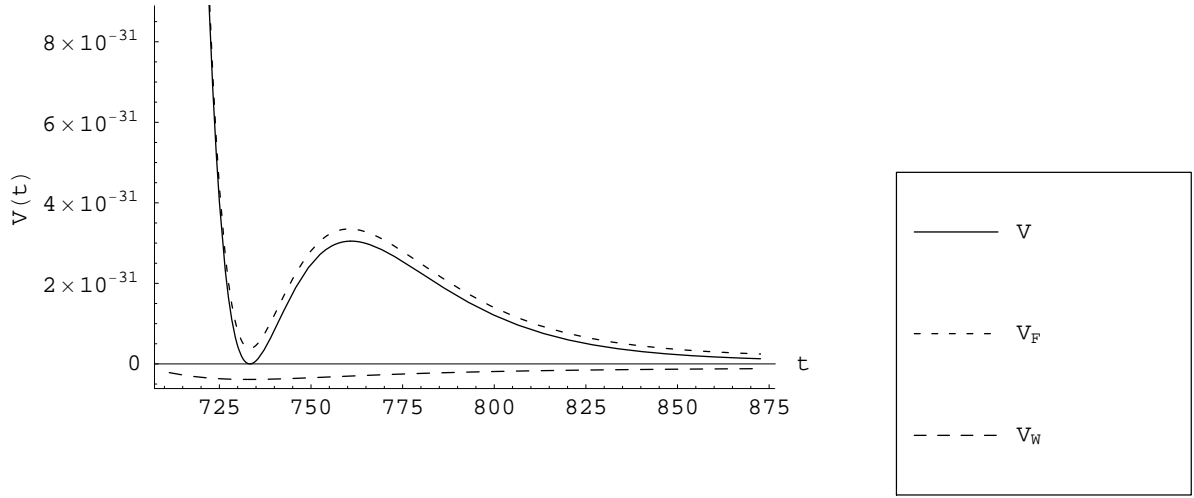}\caption{Minimisation with a cancellable D-term. The same potentials are plotted as in figure \ref{fig:noncan}, but with $V_D$ removed for clarity and again, $M_P=1$ and all other fields are at their minima.} \label{fig:can}
\end{figure}

We see that the two models, with similar input parameter values, produce remarkably similar output, but with a few important
exceptions.  Firstly, $m_{3/2}$ differs markedly between the two models due to the inflation of the potential by $e^{c^2}$.
Secondly, we see that $m_{x_2}$ is many orders of magnitude larger when the D-term is cancellable.  This appears because, despite $V_D$
and its first derivative being negligible, the second derivative of $V_D$ is naturally Planck scale.  Notice that the vev of $t$
is essentially the same in both models with the potential having a very similar form as seen in figures \ref{fig:noncan} and
\ref{fig:can}.   The shows that, since Eq.~(\ref{eq:approx_t_vev}) yields $t=735.816$ and $t=737.096$ for the non-cancellable and cancellable cases respectively, we can conclude that the approximation that the minimum is given by $F_T=0$, and hence Eq.~(\ref{eq:approx_t_vev}), is valid for these input parameters.

It is clear that, despite the differences between the models, the fine tuning of $\<V\>$ to zero can be readily achieved in both cases.
The methods of cancellation differ considerably between the two models and it is necessary to consider them separately.  Firstly
we consider the effect of the cancellable D-term.  In this case by setting $W_0=0$ and then varying $q$ we see that a small enough
charge will lift the potential to be positive.  If a small $W_0$ is then introduced, the phases will align to ensure that the
cross terms produce a negative contribution to the potential (the $W_0^2$ terms have negligible effect).  It has been demonstrated
numerically that even small values of $W_0$, such as $10^{-11}$, will produce a negative cosmological constant, assuming that $q=1/2$, while
$W_0=0$ gives $V>0$.  Hence for some value of $W_0$ in the range $0<W_{0_c}<10^{-11}$ it follows that $V=0$ for $W_0=W_{0_c}$.
When varying $W_0$ to tune $V$ to zero it is clear that the value of $m_{3/2}$ is essentially independent of the tuning.  To be
more precise, near $W_{0_c}$ the addition of a small $\Delta W_0$ will make $V<0$ whereas subtracting $\Delta W_0$ will give
$V>0$.  This will be true in the limit that $\Delta W_0 \rightarrow 0$, whereas $\Delta m_{3/2} \rightarrow 0$ in this limit.
Finally we see that because SUSY is always broken due to the effects of the meson fields $V=0$ has to be obtained by a
cancellation between $V_F$ and $V_W$.  This shows that $V=0$ implies a finite $m_{3/2}$ if $t$ stabilises.

In the simple models given here the phase of the field $C$ has no potential whatsoever. However, it is clear that one can add a nontrivial superpotential 
$\tilde{W}(C)$ to stabilise the phase without affecting stabilisation and cancellation, as long as its expectation value remains smaller than $\langle W \rangle$.  

The non-cancellable D-term lifting is both indirect and direct, with the direct lifting from the positive definite contribution to
the potential introduced by $V_D$.  The indirect effect of the non-cancellable D-term is similar to the cancellable D-term's, the
argument of the D-term is driven to zero, which necessarily moves $x_1$, $x_2$ and $t$ away from the minimum of $V_F + V_W$ and
F-term lifting occurs.  In addition to this effect the D-term provides direct lifting analogous to the original KKLT models, with
both of these effects being controlled by the size of $\delta$.  Hence varying $\delta$ allows the tuning of $V=0$.

Since, in the large $t$ limit, $V_D$ is approximately given by $\delta^2(x_1^2+x_2^2)/t$, it is clear that, in this limit,
$|\partial V_D/\partial t| \ll |V_D|$, whereas the exponential dependence of $V_F$ ensures that, in this limit, $|\partial V_F /
\partial t|\sim |\frac{2}{N_1}V_F|\sim |\frac{2}{N_2}V_F|$.  Since $|V_F|$ is of the same order as $|V_W|$ and $V$ is tuned to zero, $V_F+V_W+V_D=0$ implies that $V_D$ should be of the same order as $|V_F|$ and $|V_W|$.  This, coupled with the fact that  $\frac{2}{N_1}\gg \frac{1}{t}$, implies that  $|\partial V_F/\partial t |\gg |\partial V_D/\partial t|$. As a result the D-term does not
disturb the minimum of $t$.  The same is not true for $x_1$ and $x_2$ since $V_D$, $V_F$ and $V_W$ are all polynomial functions of
$x_1$ and $x_2$ (ignoring the $e^K$ factor which is irrelevant to this discussion) and so, very roughly\footnote{$V_F$ has a much
  larger first derivatives than $V_W$ due to the effect of the K\"ahler derivatives.}, $\partial V_F/ \partial x_i \sim V_F/x_i$,
$\partial V_D/ \partial x_i \sim V_D/x_i$ and $\partial V_W/ \partial x_i \sim 0$.  This demonstrates that $V_D$ and $V_F$ have
comparable effects on the minima of $x_i$ so as a result we expect the minima to shift upon the introduction of $V_D$.

\section{Summary and outlook}

The issue of fine-tuning necessary to achieve satisfactory degree of
hierarchy between the moduli masses, gra\-vitino mass and the scale of
cosmological constant has been revisited in the context of
supergravities with consistent D-terms. We have studied models where
supersymmetry breaking and moduli stabilisation cannot be separated
from each other: unlike in the models considered in
\cite{Kallosh:2004yh,Kallosh:2006dv} supersymmetry breaking, for finite
field vevs, is mandatory.  It turns out that even in such a case the
realistic hierarchy can be achieved on the expense of a single
fine-tuning. In models with cancellable D-terms the tuned parameter
turns out to be the constant term in the superpotential, in models
with non-cancellable D-term tuning can be restricted to the
$U(1)$-charges of moduli.  We have studied a refined mechanism of
supersymmetric F-term uplifting in the presence of consistent
D-terms. We have employed the racetrack scheme enhanced by a constant
term in the superpotential.  In contrast to the original KKLT
construction minimisation is not due to cancellation between $W_0$ and
the modulus field, but between the separate condensate terms. In
addition we introduce a spectator charged scalar. As a result, we can
easily, without any dramatic tree-level hierarchy among the
dimensionful parameters of the Lagrangian, achieve nearly vanishing
positive or negative (or zero) cosmological constant with all moduli
stabilised and small gravitino mass. The D-terms vanish at the
interesting minima, and the cancellation of the negative $-3 e^K
|W|^2$ is realised by the F-terms of the moduli fields.  Despite going
to zero in the minimum the D-term has a large effect on the mass
matrix, inducing a large splitting in the moduli mass spectrum. The
existence of a state heavier than the condensation scale is consistent
with the fact that stabilisation results from the interplay between
the gauge sector and gravity-suppressed contributions.
A distinct feature of the model discussed here is that except $W_0$,
which removes a zero eigenvalue from the mass matrix, all scales in
the hidden sector are dynamically generated via running of field
dependent gauge couplings, as implied by string theory, in contrast to
models with ISS-type hidden sectors considered earlier
\cite{Dudas:2006gr,Kallosh:2006dv}.  Perturbing the model by a small
(smaller than $<W>$) superpotential, $\tilde{W}(C)$, for the extra
charged scalar doesn't materially affect the results. Finally, the
amount of tuning necessary in the models discussed here is comparable
to those in
\cite{Dudas:2005vv,Dudas:2006vc,Dudas:2006gr,Kallosh:2006dv}.

\bigskip

\centerline{\Large \bf Acknowledgements}

\vspace*{0.5cm} 
\noindent It is a pleasure to thank Stefan Pokorski and Emilian Dudas for helpful discussions.\\
 Z.L. and O.E.-W. thank the CERN Theory Division for hospitality.  This work was partially supported by the EC 6th
Framework Programme MRTN-CT-2006-035863,
by TOK Project MTKD-CT-2005-029466, and by the Polish State Committee for Scientific Research grant KBN 1 P03D 014 26.


\vspace*{.5cm} 

\begin{thebibliography}{99}

\bibitem{Kachru:2003aw} 
S.~Kachru, R.~Kallosh, A.~Linde and S.~P.~Trivedi,
Phys.\ Rev.\ D \textbf{68} (2003) 046005 
[arXiv:hep-th/0301240]. 

\bibitem{Kallosh:2004yh}
  R.~Kallosh and A.~Linde,
  JHEP {\bf 0412} (2004) 004
  [arXiv:hep-th/0411011].

\bibitem{Achucarro:2006zf}
  A.~Achucarro, B.~de Carlos, J.~A.~Casas and L.~Doplicher,
  JHEP {\bf 0606} (2006) 014
  [arXiv:hep-th/0601190].

\bibitem{Braun:2006se}
  A.~P.~Braun, A.~Hebecker and M.~Trapletti,
  arXiv:hep-th/0611102.

\bibitem{Dudas:2005vv}
  E.~Dudas and S.~K.~Vempati,
  Nucl.\ Phys.\ B {\bf 727} (2005) 139
  [arXiv:hep-th/0506172].


\bibitem{Dudas:2006vc}
  E.~Dudas and Y.~Mambrini,
  JHEP {\bf 0610} (2006) 044
  [arXiv:hep-th/0607077].

\bibitem{Choi:2006bh}
  K.~Choi and K.~S.~Jeong,
  JHEP {\bf 0608} (2006) 007
  [arXiv:hep-th/0605108].

\bibitem{Villadoro:2005yq}
  G.~Villadoro and F.~Zwirner,
  Phys.\ Rev.\ Lett.\  {\bf 95}, 231602 (2005)
  [arXiv:hep-th/0508167].

\bibitem{Villadoro:2006ia}
  G.~Villadoro and F.~Zwirner,
  JHEP {\bf 0603} (2006) 087
  [arXiv:hep-th/0602120].


\bibitem{Lalak:2005hr}
  Z.~Lalak, G.~G.~Ross and S.~Sarkar,
  Nucl.\ Phys.\ B (in press) [arXiv:hep-th/0503178].

\bibitem{Ellis:2006ar}
  J.~Ellis, Z.~Lalak, S.~Pokorski and K.~Turzynski,
  JCAP {\bf 0610}, 005 (2006)
  [arXiv:hep-th/0606133].

\bibitem{Gomez-Reino:2006dk}
  M.~Gomez-Reino and C.~A.~Scrucca,
  JHEP {\bf 0605}, 015 (2006)
  [arXiv:hep-th/0602246].

\bibitem{Lebedev:2006qq}
  O.~Lebedev, H.~P.~Nilles and M.~Ratz,
  Phys.\ Lett.\ B {\bf 636} (2006) 126
  [arXiv:hep-th/0603047].

\bibitem{Intriligator:2006dd}
  K.~Intriligator, N.~Seiberg and D.~Shih,
  JHEP {\bf 0604} (2006) 021
  [arXiv:hep-th/0602239].


\bibitem{Dudas:2006gr}
  E.~Dudas, C.~Papineau and S.~Pokorski,
  [arXiv:hep-th/0610297].


\bibitem{Kallosh:2006dv}
  R.~Kallosh and A.~Linde,
  [arXiv:hep-th/0611183].


\bibitem{Abe:2006xp}
  H.~Abe, T.~Higaki, T.~Kobayashi and Y.~Omura,
  arXiv:hep-th/0611024.

\bibitem{Brummer:2006dg}
  F.~Brummer, A.~Hebecker and M.~Trapletti,
  Nucl.\ Phys.\ B {\bf 755} (2006) 186
  [arXiv:hep-th/0605232].

\bibitem{Choi:2004sx}
  K.~Choi, A.~Falkowski, H.~P.~Nilles, M.~Olechowski and S.~Pokorski,
  JHEP {\bf 0411} (2004) 076
  [arXiv:hep-th/0411066].

\bibitem{Burgess:2003ic}
  C.~P.~Burgess, R.~Kallosh and F.~Quevedo,
  JHEP {\bf 0310} (2003) 056
  [arXiv:hep-th/0309187].




\bibitem{Binetruy:2004hh}
  P.~Binetruy, G.~Dvali, R.~Kallosh and A.~Van Proeyen,
  Class.\ Quant.\ Grav.\  {\bf 21} (2004) 3137
  [arXiv:hep-th/0402046].

\bibitem{Haack:2006cy}
  M.~Haack, D.~Krefl, D.~Lust, A.~Van Proeyen and M.~Zagermann,
  JHEP {\bf 0701}, 078 (2007)
  [arXiv:hep-th/0609211].


\end{thebibliography}
\end{document}